\documentclass[pra,showpacs,twocolumn,eqsecnum,superscriptaddress]{revtex4}
\usepackage{makeidx}
\usepackage{eurosym}
\usepackage{amssymb}
\usepackage{amsfonts}
\usepackage{amsmath}
\usepackage{setspace}
\usepackage[english]{babel}
\usepackage{graphicx}
\usepackage[applemac]{inputenc}
\usepackage{color}
\usepackage{ifthen}
\usepackage{float}
\usepackage{verbatim}
\usepackage{subfigure}

\begin{document}
\title{Bose-Einstein Condensates in a Cavity-mediated Triple-well}

\author{Lei Tan}
 \affiliation{Institute of Theoretical Physics, Lanzhou University, Lanzhou $730000$,
China} \affiliation{Department of Physics and Astronomy,
University College London, Gower Street, London WC1E 6BT,
United Kingdom} \affiliation{Beijing National Laboratory for Condensed Matter
Physics, Institute of Physics, Chinese Academy of Sciences,
Beijing 100190, China}

\author{Bin Wang}
\affiliation{Institute of Theoretical Physics, Lanzhou
University, Lanzhou $730000$, China}

\author{Peter Barker}
\affiliation{Department of Physics and Astronomy, University
College London, Gower Street, London WC1E 6BT, United Kingdom}
\date{\today}

\author{Wu-Ming Liu}
\affiliation{Beijing National Laboratory for Condensed Matter
Physics, Institute of Physics, Chinese Academy of Sciences,
Beijing 100190, China}

\begin{abstract}
We investigate the energy structures and the dynamics of a Bose-Einstein condensates (BEC) in a triple-well potential coupled a high finesse optical cavity within a mean field approach. Due to the intrinsic atom-cavity field nonlinearity, several interesting phenomena arise which are the focuses of this work.  For the energy structure, the bistability appears in the energy levels due to this atoms-cavity field nonlinearity, and the same phenomena can be found in the intra-cavity photons number. With an increase of the pump-cavity detunings, the higher and lower energy levels show a loop structure due to this cavity-mediated effects. In the dynamical process, an extensive numerical simulation of localization of the BECs for atoms initially trapped in one-, two-, and three-wells are performed for the symmetric and asymmetric cases in detail. It is shown that the the transition from oscillation to the localization can be modified by the cavity-mediated potential, which will enlarge the regions of oscillation. With the increasing of the atomic interaction, the oscillation is blocked and the localization emerges. The condensates atoms can be trapped either in one-, two-, or in three wells eventually where they are initially uploaded for certain parameters.  In particular, we find that the transition from the oscillation to the localization is accompanied with some irregular regime where tunneling dynamics is dominated by chaos for this cavity-mediated system.
\end{abstract}
\pacs{42.50.Pq, 03.75.Lm, 05.45.-a}
\maketitle
\section{Introduction}
Cavity-mediated Bose-Einstein condensates system (BECs) has been widely investigated as a promising platform to explore the exotic many-body phenomena in recent years both in theory and experiment due to the controllability of both the confining potential geometry and the interatomic interaction \cite{Brennecke,Colombe}. If the cavity resonance and the atomic resonance are in the large-detuning limit, the dispersive regime can be realized. At this stage, the strong coupling between the atoms and the cavity field will induces an additional optical potential for the atoms and exerts significant mechanical forces on the motion of the atoms; meanwhile, the atoms imprint a position dependence phase shift on the cavity field vice versa. As the field adjusts accordingly, the field causes a back action on the atoms, resulting in the nonlinear atom-field interplay \cite{Cola,Slama,Vukics,Maschler,Zhang,Larson,Szirmai}. This highly nonlocal nonlinearity is quite different from the usual local atom-atom interactions. As the cavity-atom interaction induces an optical lattice for the atoms, then the cavity-mediated optical lattice model is one of the usual platform for discussing the cavity-mediated ultracold atoms and BECs \cite{Ruostekoski}. Based on this nonlinear interactions, the cavity field coupling either to the center-of-mass \cite{Murch,Brennecke1,Baumann} or to the spin degrees of freedom \cite{Zhou} and to the both cases \cite{Larson1,Dong} i.e., the nonlinear coupling among the external and internal states are demonstrated.  Recently, rapid proposals have been intensely investigated, which gives rise to a number of interesting nonclassical features such as optical bistability \cite{Gupta,Ritter,Szirmai1,Yang,Prasanna,Zhou1}, collective dynamics \cite{Keeling} and quantum phase transitions \cite{Nagy,Liu}.

On one hand, a cavity-mediated double-well potential is the simplest case of a multiwell potential for studying the cavity-mediated BECs system, a plethora of novel phenomena, such as strong population transfer, the complex and diverse stationary behaviors and the pseudo-self-trapping effect, have been observed and analyzed in great detail for the this system because of the unique cavity-mediated nonlinearity in the context of cavity quantum electrodynamics \cite{Zhang,Zhang1,Larson2}.  The role of dissipation in the system of a light field coupled with ultracold atoms trapped in a two-well optical potential in a lossy cavity is also investigated \cite{Chen}. Special asymmetric double-well case is studied too, the results show that the performed numerical simulations on bistability in optical lattice have good agreement with the special asymmetric double-well case described by a two-mode model \cite{Zhou1}.

On the other hand, If one wants to make a extension of the double-well system, the triple well is the straightforward natural one. Compared with the double well system, the difference is essential both in mathematics and physics for the triple well system, which has the following distinguished characteristics.  A triple well potential has a much richer structure \cite{Mossmann} and can supports the simplest model for exhibiting the effect of next-nearest-neighbor coupling on quantum tunneling \cite{Lahaye}, the quantum tunneling may happen between several wells simultaneously, then the tunneling dynamics in the triple well will show more interesting behavior \cite{Liu1}. The study of the triple well system will also provide a bridge between the simple double well and the multiwell systems, as a protype of optical lattices, helping us provide a bottom-up understanding of mechanisms operating in the infinite optical lattice \cite{Cao}. Moreover, similarly to the analogue of the double well to the Josephson junction, the triple well is the minimum system that can model the source-gate-drain junction, and draws much attention from the perspective of atomtronics. Considering these reasons, recently, then there is a growing substantial attention in the dynamics of ultracold atomic clouds in triple-well potentials, A variety of proposals to achieve controllable atomic dynamics based on the triple well have been presented extensively by employing the standard models of the Gross-Pitaevskii
equation and the Bose-Hubbard Hamiltonian mainly. It includes the eigenstates and the tunneling dynamics \cite{Cao,Graefe,Lu,Streltsov}, the transistorlike behavior \cite{Stickney}, the adiabatic transport of a hole \cite{Benseny}, and the mesoscopic quantum superpositions \cite{Lahaye}. However, unlike the optical lattice and double well cases, no work has been studied in the triple well coupled to an optical cavity so far and a fundamental question of whether or not loop structure and chaotic exist is still open. Then theoretical predictions on the dynamics of cavity-mediated BECs system in a triple well are highly desirable. On the contrary, as we will show below, the cavity-mediated BECs system does introduce qualitatively novel physics in triple well systems. We will discuss how the nonlinear interaction leads to s loop structure in the energy band. In addition, we also show how this nonlocal nonlinearity leads to a transition quantum dynamics characterized by striking new phenomena, including the chaotic quantum dynamics.

This paper is organized as follows. In Sec. II the mean-field approximation is employed to map the quantum model into a classical Hamiltonian describing the atoms and the corresponding dynamical equations are derived. In Sec. III we study the optical bistability in the atom-cavity system and achieve how the intra-cavity-photons and energy band structure depends on the cavity detunings. In Sec. IV we investigate the transition to self-trapping of BECs in one well and in two wells, respectively, showing some interesting regions that the chaotic motion and steady behavior emerge alternately. The topics are extended to a tilted triple-well system in Sec. V. Finally, the results are briefly summarized in Section VI.

\section{the basic model}

Consider a BECs system confined to in a triple well coupled to an external single mode of high-finesse optical cavity (schematically sketched as in Fig.(1)).  This system may be constructed by splitting a BECs, which is already coupled to a cavity mode, into three weakly linked condensates in many ways. In the large-detuning limit and in the rotating frame at the pump frequency, the Hamiltonian for the condensate system
can be written as ($\hbar$=1 throughout),
\begin{eqnarray}
H=H_{a}+H_{f}+H_{int},
\label{eq-H}
\end{eqnarray}
$H_a$ is the condensates Hamiltonian in the three-mode approximation,
\begin{eqnarray}
 H_{a}=&\lambda(b_{1}^{\dag}b_{1}-b_{3}^{\dag}b_{3})
-v(b_{1}^{\dag}b_{2}+b_{2}^{\dag}b_{1}+b_{3}^{\dag}b_{2}+b_{2}^{\dag}b_{3})\nonumber\\
&+\frac{1}{2}c(b_{1}^{\dag}b_{1}^{\dag}b_1b_1+b_{2}^{\dag}b_{2}^{\dag}b_2b_2+b_{3}^{\dag}b_{3}^{\dag}b_3b_3),
 \end{eqnarray}

\begin{figure}[ptb]
\centerline{\includegraphics*[width=0.5\textwidth]{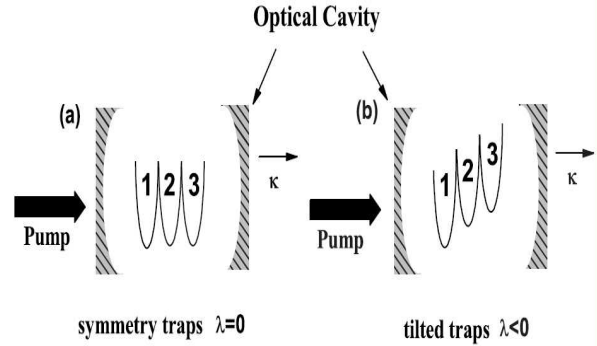}}\caption{
The sketch of cavity-mediated triple-well system in which (a) is the symmetry case ($\lambda=0$) and
(b) is the asymmetry case ($\lambda<0$). $\lambda$ is the zero-point energy in each well and the BECs will be trapped in the wells. The cavity is
driven by a pump field with an amplitude $\eta$ and its loss rate is $\kappa$.}%
\label{Fig1}%
\end{figure}

Where $b_i^{\dag}(b_i)(i=1,2,3)$ creates (annihilates) an atom in its internal ground state in the $ith$ traps, $v$ is the tunneling matrix element between the nearest modes, while $c$ denotes the repulsive interaction strength between a pair of atoms in the same mode. $\lambda$ is the zero-energy of the wells. $H_f$ is the external single-mode field Hamiltonian,
\begin{eqnarray}
H_{f}=\omega_{c}a^{\dag}a+\eta(t){\rm{e}}^{{\rm{-i}}\omega_{p}t}a^{\dag}+\eta^{*}(t){\rm{e}}^{{\rm{i}}\omega_{p}t}a,
\end{eqnarray}
where $\omega_c$ and $\omega_p$ are the cavity-mode frequency and the pump frequency, respectively. $\eta(t)$ is the amplitude of the pump field and varies slowly, $\rm{i.e.,}$  $|\dot{\eta}/\eta|\ll\omega_p$. In the limit of large detuning \cite{Colombe} and the weak pump field, the atom-cavity field interaction is of dispersive nature, in this sense the upper level of the atoms can be adiabatically eliminated. Under the three-mode approximation for the BECs and dropping the coupling terms between the cavity mode and the atomic tunneling, the interaction Hamiltonian between BECs and the cavity mode can be obtained,
\begin{eqnarray}
H_{int}=U_{0}a^{\dag}a(J_{1}b_1^{\dag}b_1+J_{2}b_2^{\dag}b_2+J_{3}b_3^{\dag}b_3).\label{Hint}
\end{eqnarray}
where $U_0={g_0}^{2}/(\omega_c-\omega_a)$ characterizes the strength of the light shift per photon that an atom may experience, with $g_0$ being the atom-cavity mode coupling constant. The parameter $J_i( i=1,2,3)$ denotes the overlap between the atomic mode and the cavity mode. From Eq. (\ref{Hint}), one may immediately find that the frequency cavity mode is renormalized due to the atom-field interaction, meanwhile the atom ensemble feels a induced potential. This atom-field nonlinearity will induce novel phenomena in the triple well system. According to the Heisenberg's equation, a dimensionless Schr\"{o}dinger equation of this cavity-mediated triple-well system can be written as
\begin{eqnarray}
&{\rm{i}}\dot{b_1}&=-vb_2+\lambda b_1+cb_1^{\dag}b_1b_1+J_1U_0a^{\dag}ab_1\\\nonumber
&{\rm{i}}\dot{b_2}&=-vb_1+vb_3+cb_2^{\dag}b_2b_2+J_2U_0a^{\dag}ab_2\\\label{eq-1}
&{\rm{i}}\dot{b_3}&=-vb_2-\lambda b_3+cb_3^{\dag}b_3b_3+J_3U_0a^{\dag}ab_3\nonumber
\end{eqnarray}
\begin{eqnarray}
{\rm{i}}\dot{a}=&[\omega_c+U_0(J_1b_1^{\dag}b_1+J_2b_2^{\dag}b_2+J_3b_3^{\dag}b_3)]a\nonumber\\
&-{\rm{i}}\kappa a+\eta(t)e^{{\rm{-i}}\omega_pt},
\label{eq-2}
 \end{eqnarray}
the cavity loss $\kappa$ is induced phenomenologically, which is the dominant dissipation process since spontaneous emission is suppressed under the large detuning approximation. Eq.(2.5) shows that, the dispersive interaction between cavity photons and the condensate atoms introduces an effective light shift to the states $b_i(i=1,2,3)$.  The characteristic values of $\kappa$ of the high-finesse optical cavity, the order of $2\pi\times 10^6$ Hz, is typically much larger than the spin oscillation frequency with the order of $(2\pi\times 10)-(2\pi\times 10^2)$ Hz,  This means that the cavity mode can follow the condensates wave functions adiabatically. Taking the standard mean-field approximation, we  treat the operators $b_i(i=1,2,3)$ and $a$ as classical quantities, $b_1=\sqrt{N_1}e^{\rm{i}\theta_1}$, $b_2=\sqrt{N_2}e^{\rm{i}\theta_2}$, $b_3=\sqrt{N_3}e^{\rm{i}\theta_3}$, and $a\sim \alpha$. Here $N_i(i=1,2,3)$ are the numbers of atoms in the three wells, and $\theta_i(i=1,2,3)$ are their phases. The relative phases are $\phi_1=\theta_1-\theta_2$ and $\phi_3=\theta_3-\theta_2$. And the total number of atoms $N=N_1+N_2+N_3$. Then the intracavity field amplitude can be derived from Eq.(2.6)
\begin{eqnarray}
\langle a\rangle\equiv\alpha(t)=
\frac{\eta(t)e^{{\rm{-i}}\omega_pt}}{{\rm{-i}}\kappa+[\omega_p-\omega_c-U_0(J_1N_1+J_2N_2+J_3N_3)]}
 \label{eq-5}
\end{eqnarray}
and the intracavity photon number is
\begin{eqnarray}
\langle a^{\dag}a \rangle\equiv|\alpha(t)|^{2}=\frac{|\eta(t)|^2}{\kappa^2+[\Delta-J_2NU_0+\delta U_0(N_1+N_3)]^2}.
\label{eq-6}
\end{eqnarray}
where $\Delta\equiv\omega_p-\omega_c-J_2NU_0$ and $\delta\equiv J_2-J_1$
$(\equiv J_2-J_3)$ is the coupling difference between the two atomic
modes to the same cavity mode. From Eq. (\ref{eq-6}), one may easily conclude that, the important feature of the cavity-induced effective light shift is that it is sensitive to the population distribution of the condensates in the triple well, which in turn will induce novel nonlinear phenomena in BECs.  By introducing three dimensionless parameters $A(t)=\eta(t)/[\delta NU_0]$, $B=\Delta/[\delta NU_0]$, and $C=\kappa/[\delta NU_0]$ named the reduced pumping strength, the reduced detuning, and the reduced loss rate, respectively. The photon number Eq. (\ref{eq-6}) can be written as
\begin{eqnarray}
|\alpha(t)|^{2}=\frac{A(t)^2}{[n_1+n_3+B]^2+C^2}
\end{eqnarray}
where $n_i$=$N_i/N(i=1,2,3)$,  and $n_1+n_2+n_3=1$. Then the Eqs.~(5) can be rewritten in terms of $n_i(i=1,3)$ and the phase difference $\phi_i(i=1,3)$ as

\begin{figure}[ptb]
\centerline{\includegraphics*[width=0.5\textwidth]{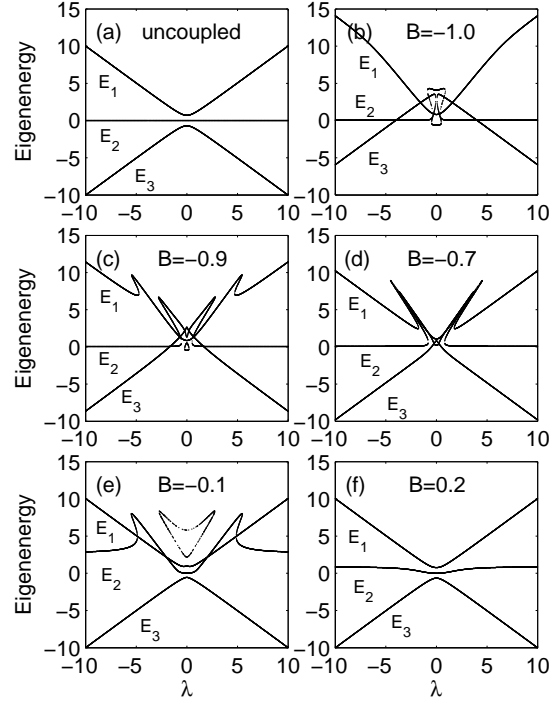}}\caption{The eigenenergy band structures of the condensates both for
the uncoupled (a) and coupled (b-f) cases. From (b) to (f), the reduced cavity detuning $B$ is varied. The other parmeters are set as $\delta U_0A^2/(2v)=0.02$, $C=0.07$.}%
\label{Fig2}%
\end{figure}

\begin{figure}[ptb]
\centerline{\includegraphics*[width=0.5\textwidth]{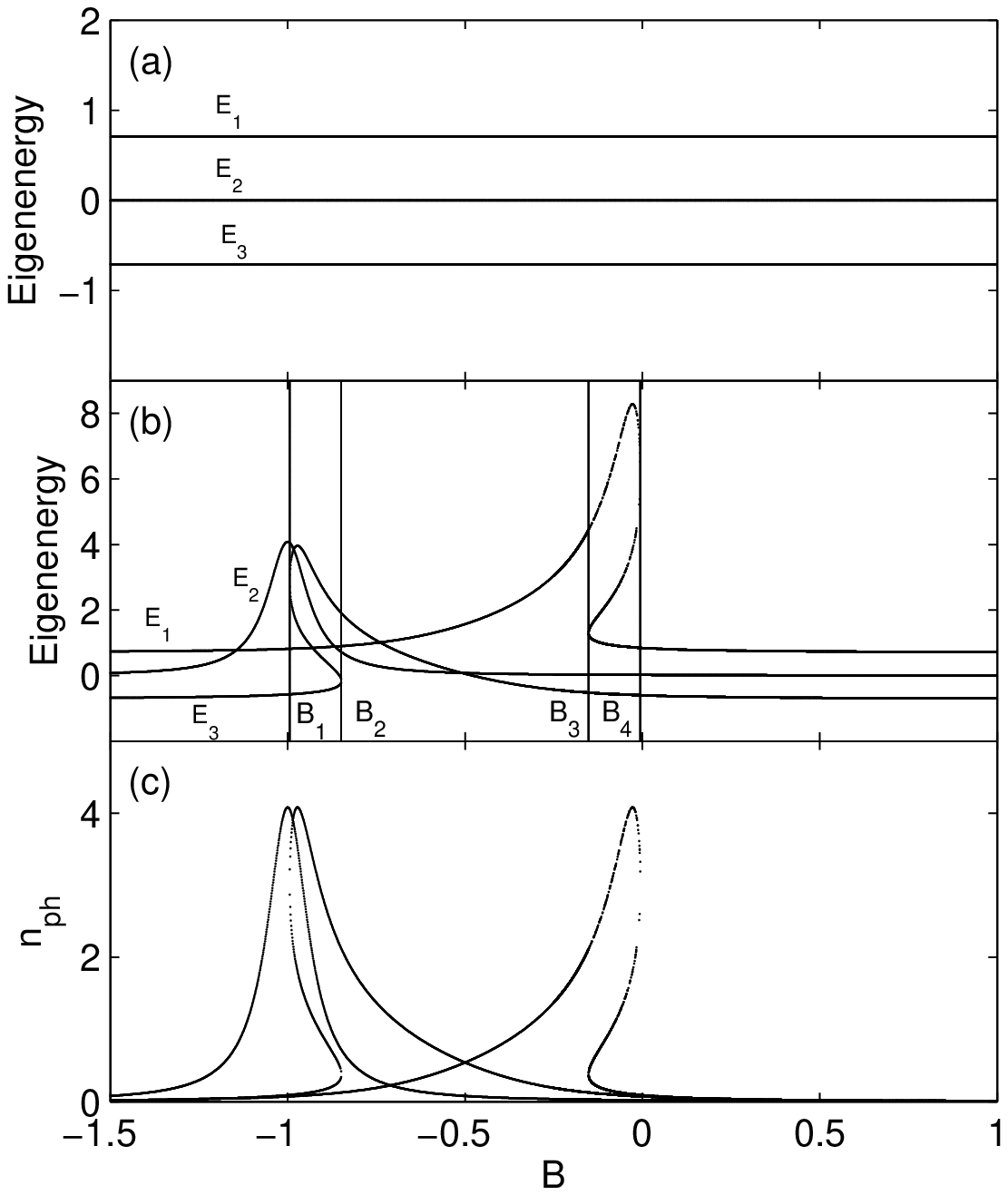}}\caption{For the parameters $\delta U_0A^2/(2v)=0.02$, $C=0.07$, $r=0$, $\lambda=0$,
the eigenenergy levels of the condensates (a) uncoupled and (b) coupled to a
single-cavity mode given by Eq.(\ref{eq-4}) and (c) the intra cavity photon number
$|\alpha|^2$ versus the reduced cavity detuning $B$.}%
\label{Fig3}%
\end{figure}

\begin{eqnarray}
\dot{n_1}=\sqrt{n_1}\sqrt{1-n_1-n_3}\sin{\phi_1},
\label{eq-7}
\end{eqnarray}
\begin{eqnarray}
\dot{n_3}=\sqrt{n_3}\sqrt{1-n_1-n_3}\sin{\phi_3},
\label{eq-8}
\end{eqnarray}
\begin{eqnarray}
\dot{\phi_1}=&-\lambda-r[n_1-(1-n_1-n_3)]+\frac{1}{2}\frac{\sqrt{1-n_1-n_3}}{\sqrt{n_1}}\cos{\phi_1}\nonumber\\
             &-\frac{1}{2}\frac{(\sqrt{n_1}\cos{\phi_1}+\sqrt{n_3}\cos{\phi_3})}{\sqrt{1-n_1-n_3}}+\frac{\delta U_0}{2v}|\alpha|^{2},
\label{eq-9}
\end{eqnarray}
\begin{eqnarray}
\dot{\phi_3}=&\lambda-r[n_3-(1-n_1-n_3)]+\frac{1}{2}\frac{\sqrt{1-n_1-n_3}}{\sqrt{n_3}}\cos{\phi_3}\nonumber\\
            &-\frac{1}{2}\frac{(\sqrt{n_1}\cos{\phi_1}+\sqrt{n_3}\cos{\phi_3})}{\sqrt{1-n_1-n_3}}+\frac{\delta U_0}{2v}|\alpha|^{2},
\label{eq-10}
\end{eqnarray}
where the time is rescaled in units of the Rabi oscillation time 1/(2$v$), $2vt\rightarrow t$. The dimensionless parameter $r\equiv Nc/(2v)$ measures the interaction strength against the tunneling strength.  From Eqs. (\ref{eq-9}) and (\ref{eq-10}), one can find that the cavity field influences the atomic dynamics in the same manner as the the zero-energy, i.e.$\lambda$, of the triple wells system, which will lead to a redistribution of the atomic population among different states $b_i$ based on the particle tunneling. Importantly, this cavity-induced effective zero-energy is dependent on the atomic population distribution, as manifested by Eq. (\ref{eq-6}). It is precisely this interdependence of the atomic dynamics and the intracavity photon number that will result in some interesting nonlinear dynamics of this cavity-mediated system, which is the focus of this paper. Furthermore, a classical Hamiltonian $H_c(n_1,n_3,\phi_1,\phi_3,t)$ can be constructed, in terms of two pairs of conjugate variables $n_1$ and $\phi_1$, $n_3$ and $\phi_3$, using $\dot{n_i}=-\frac{\partial H_c}{\partial \phi_i}$, $\dot{\phi_i}=\frac{\partial H_c}{\partial n_i}$. Such a counterpart mean-field Hamiltonian of the quantum Hamiltonian (\ref{eq-H}) reads
\begin{eqnarray}
H_c=&\lambda(n_3-n_1)-\frac{1}{2}r[n_1^{2}+n_3^{2}+{(1-n_1-n_3)}^{2}]\nonumber\\
    &+\sqrt{1-n_1-n_3}(\sqrt{n_1}\cos{\phi_1}+\sqrt{n_3}\cos{\phi_3})\nonumber\\
    &-\frac{\delta U_0}{2v}F(n_1,n_3,t),
\label{eq-4}
\end{eqnarray}
with
\begin{eqnarray}
F(n_1,n_3,t)=\frac{A^2(t)}{C}\arctan[\frac{n_1+n_3+B}{C}].
\end{eqnarray}
The first three terms in Eq. (\ref{eq-4}) are the Hamiltonian of a bare Josephson Hamiltonian as first derived in Ref. \cite{Liu1}. They describe the energy cost due to the zero-energy, the phase twisting between the two condensates and the atom-atom repulsion, respectively.  The last term in Eq. (\ref{eq-4}) is the cavity-mediated interaction. In its nature, this term is similar to the potential felt as the atom passing trough the cavity field adiabatically \cite{Haroche}, which tilts the symmetric well.  In this paper, we only concentrate on the case that the pump strength is a constant, so that the cavity-mediated triple well system is autonomous and the Hamiltonian (\ref{eq-4}) is conserved in time.

\begin{figure}[ptb]
\centerline{\includegraphics*[width=0.5\textwidth]{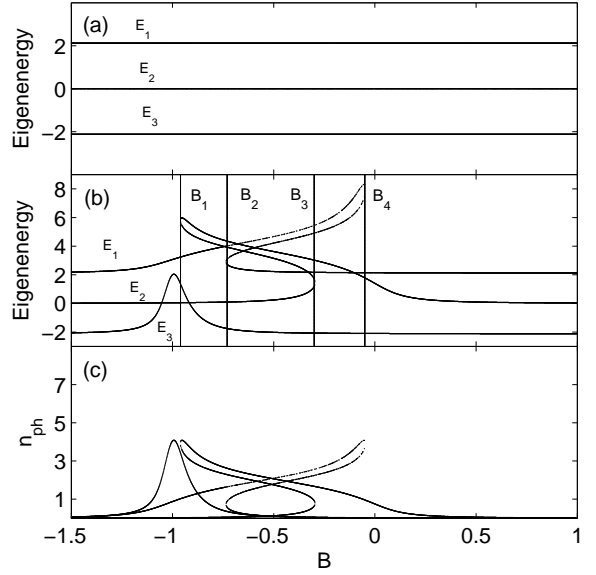}}\caption{
For the parameters $\delta U_0A^2/(2v)=0.02$, $C=0.07$, $r=0$, $\lambda=2$,
the eigenenergy levels of the condensates (a) uncoupled and (b) coupled to a
single-cavity mode given by Eq.(\ref{eq-4}) and (c) the intra cavity photon number
$|\alpha|^2$ versus the reduced cavity detuning $B$.}%
\label{Fig4}%
\end{figure}

\begin{figure}[ptb]
\centerline{\includegraphics*[width=0.5\textwidth]{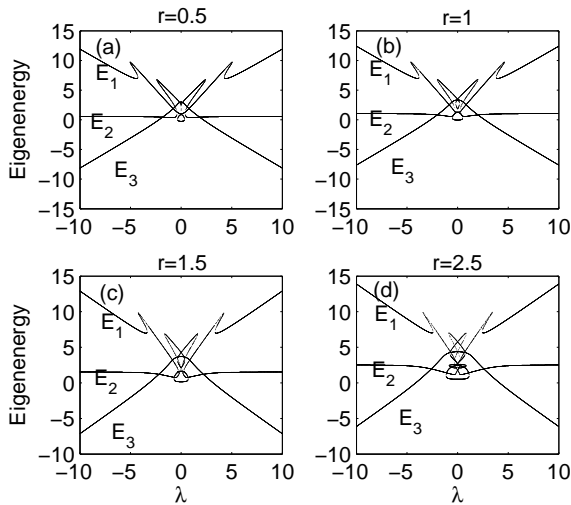}}\caption{
The eigenenergy levels as the parameter $r$ is varied
for $\delta U_0A^2/(2v)=0.02$, $B=-0.9$, $C=0.07$.}%
\label{Fig5}%
\end{figure}

\begin{figure}[ptb]
\centerline{\includegraphics*[width=0.5\textwidth]{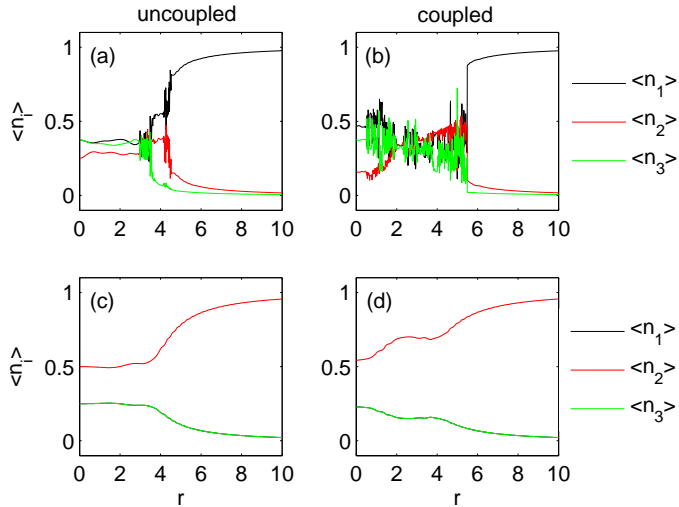}}\caption{
(Color online) The average of $\langle n_i \rangle$ $(i=1,2,3)$ of the BECs for the different interaction $r$, with initial value $n_1(0)=1$[(a) and (b)], $n_2(0)=1$[(c) and (d)] in the symmetric wells $\lambda=0$, respectively. (a) and (c) are for the uncoupled cases while (b) and (d) are for the coupled cases. The other parameters $\delta U_0A^2/(2v)=0.02$, $B=-0.9$ and $C=0.07$}%
\label{Fig6}%
\end{figure}

\section{The Energy Structure}
In this section, we will study the
eigenenergies of the condensates in the parameter regime
that supports the existence and the conditions for loop structure emergence of the adiabatic lower and higher energy levels in the cavity-mediated triple well system and presents the unique features which make them different from those in the optical lattices and double well. In order to obtain the effects of the nonlinearity to be clearly identified, we chose $c=0$ firstly, which can be achieved via Feshbach resonances or is approximately valid in a sufficiently dilute gas.  The cases including the atomic collisions are also calculated in the last part of this section for comparison with the former works \cite{Liu1}. The fixed point or minimum energy point of the classical Hamiltonian system (\ref{eq-4}) corresponds to the eigenstate of the quantum system. Solving the eigenvalue equation, one can derive the analytical expressions for these fixed points, however, the analytical solutions are lengthy, it is difficult to interpret them, then we restore to the numerical and graphic method to show the new physical effects in them \cite{Parameters}. The eigenenergies of the BECs uncoupled or coupled to the cavity field as a function of the zero-point energy bias $\lambda$ are shown in Fig. \ref{Fig2}, with Fig.2(a) for the uncoupled cases, Figs.2(b)-2(f) for the coupled cases with the increasing detunings. It can be found that the cavity-mediated BECs has a more complex and diverse energy structure. To be special, the coupled cases show the bistability behavior and the loop structure appears, which are not observed in the uncoupled counterpart, which implies that some new stationary points appear in the eigenenergies due to the cavity-mediated interaction. We can see that with the increase of the detunings, the lower energies $E_3$ and $E_2$ show bistability firstly,  then the higher energy $E_1$ does too. When the detunings are very large, the bistability behavior disappears for all the energy levels.  For the suitable detuings, the lower and the higher energies $E_3$ and $E_1$ show loop structures, which shrinks in size and vanishes as the detuning is further increased. But the energy $E_2$ does not show loop structure for the parameters considered in Fig. (\ref{Fig2}). Actually, numerical calculation for a wide parameters shows that the energy $E_2$ does not has a loop structure.  Seen from Fig. 2(b)-2(f), it can be found that the emergence condition for the bistability are different for different energies. $E_3$ only shows bistability phenomena at near $\lambda=0$ as the detunings is increased, and loop structure disappears for large detunings. $E_2$ show bistability behavior at near $\lambda=0$  when the detuning is relatively small, while for the large detunings, the bistability appears for large $\lambda$. As for $E_1$, bistability begin to appear at the regions of larger $\lambda$ for the large detunings; With the increase of the large detuning,  the bistability regimes move towards to the directions of the small zero-energy $\lambda$.
\begin{figure}[ptb]
\centerline{\includegraphics*[width=0.5\textwidth]{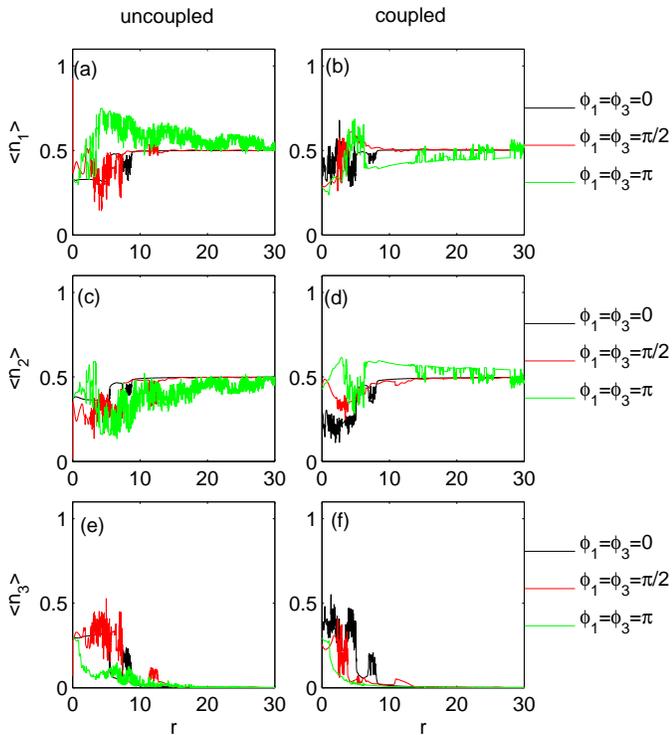}}\caption{(Color online) The average population of (a), (c), (e) (uncoupled case) and (b), (d), (f)(coupled case) $\langle n_i \rangle$  with the initial value $n_1=0.5$, $n_2=0.5$ and $n_3=0$ in the symmetric wells $\lambda=0$.  The other parameters of the cavity are chosen as $\delta U_0A^2/(2v)=0.02$, $B=-0.9$ and $C=0.07$.}%
\label{Fig7}%
\end{figure}

\begin{figure}[ptb]
\centerline{\includegraphics*[width=0.5\textwidth]{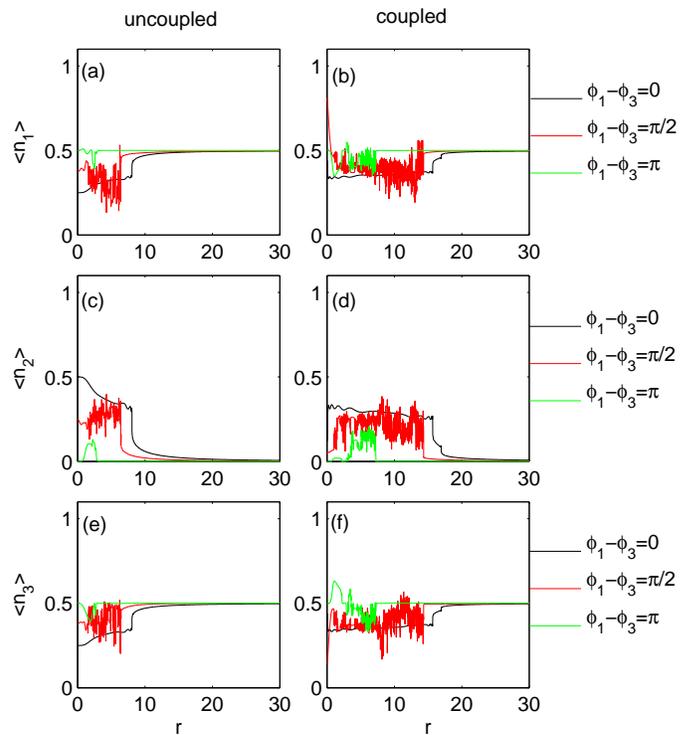}}\caption{
(Color online) The average population of (a), (c), (e) (uncoupled case) and (b), (d), (f)(coupled case) $\langle n_i \rangle$  with the initial value $n_1=0.5$, $n_3=0.5$ and $n_2=0$ in the symmetric wells $\lambda=0$. The other parameters of the cavity are chosen as $\delta U_0A^2/(2v)=0.02$, $B=-0.9$ and $C=0.07$.}%
\label{Fig8}%
\end{figure}

Looped structures have been discussed before for BECs in the ordinary (noncavity)  \cite{Wu,Diakonov,Watanabe}, the cavity-mediated optical lattices \cite{Prasanna} and the bare tripe well \cite{Liu1}, respectively,  which is important due to the possible wide implications for various systems in condensed matter physics and nuclear physics \cite{Watanabe1}. But the loop structure in Fig. (\ref{Fig2}) is qualitatively different from the ones, resulting from the interatomic interactions nonlinearity obtained in Ref. \cite{Liu1}. Here the bistability comes from the cavity-induced nonlinearity. It can be understood as follows: Because of the sensitive dependence of the BEC dynamics on the  cavity induced potential. The cavity-induced phase shift can effectively result in population redistribution among the three wells even at the single-photon level. The BEC atoms in turn collectively act as a dispersive medium shifting the cavity resonance. Bistability results from this nonlinear feedback between the photons and the atoms in different wells. With the increase of the detunings, the cavity induced potential begin to enhance the triple-well asymmetry felt by the atoms. As a result, the atoms tend to stay in the deep well. This will give rise to a larger population imbalance and shift the cavity resonance, and will again lead to the increase of the probe mode intensity, as can be seen from Eq. (7) by setting detunings$<-U_0(J_i,N_i)$.

\begin{figure}[ptb]
\centerline{\includegraphics*[width=0.5\textwidth]{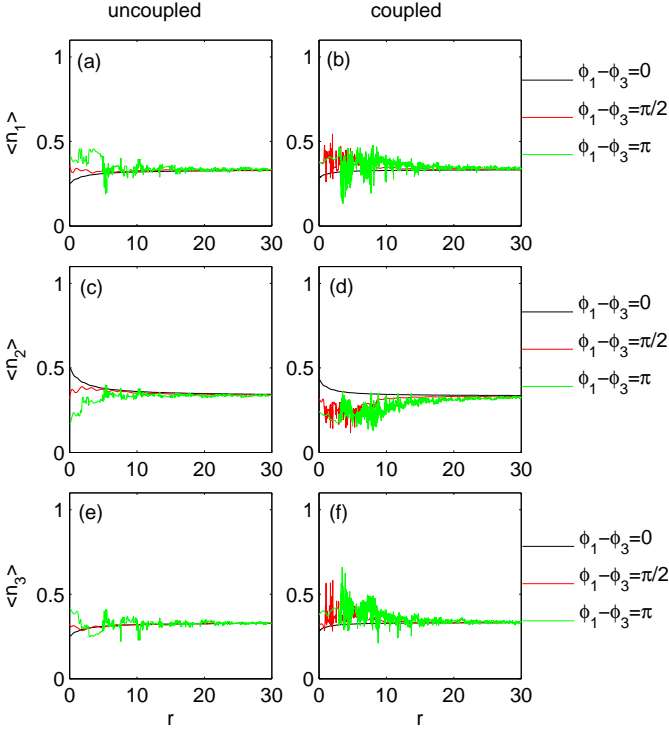}}\caption{
(Color online) The average population of (a), (c), (e) (uncoupled case) and (b), (d), (f)(coupled case) $\langle n_i \rangle$  with the initial value $n_1=1/3$, $n_2=1/3$, $n_3=1/3$ for different relative phases in the symmetric wells $\lambda=0$. The other parameters of the cavity are chosen as $\delta U_0A^2/(2v)=0.02$, $B=-0.9$ and $C=0.07$.}%
\label{Fig9}%
\end{figure}

\begin{figure}[ptb]
\centerline{\includegraphics*[width=0.5\textwidth]{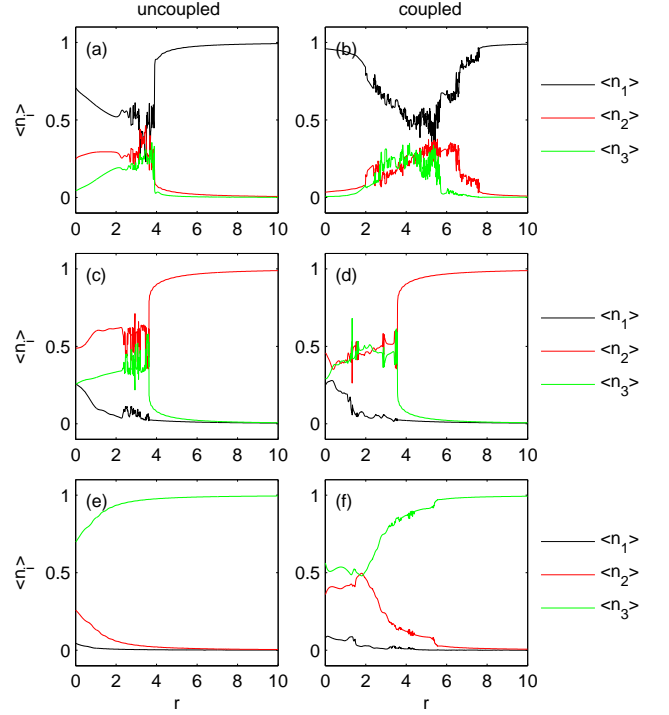}}\caption{(Color online) The average of $\langle n_i \rangle$ $(i=1,2,3)$ of the BECs for the different interaction $r$ in the tilted wells $\lambda=-1$, with initial value $n_1(0)=1$[(a) and (b)], $n_2(0)=1$[(c) and (d)], and $n_3(0)=1$[(e) and (f)] respectively. (a), (c) and (e) are for the uncoupled cases while (b), (d) and (f) are for the coupled cases. The parameters of the cavity are $\delta U_0A^2/(2v)=0.02$,
$B=-0.9$ and $C=0.07$.}%
\label{Fig10}%
\end{figure}
The trend of the energy structures and intra-cavity photons with respect to the reduced pump-cavity detunings are shown in Fig. \ref{Fig3}. We focus on the condition $\lambda=0$ and $\lambda=2$, respectively. Beginning from $\lambda=0$ case, it can be found that, the level structure is similar to the uncoupled counterpart for the small detuning ($B<B_1=-0.994$). While with an increasing detunings ($B_1<B<B_2=-0.849$), the bistable curve appears in the lower energy level, which corresponds to the loop structure of the lower level $E_3$ in Fig. \ref{Fig2}(c). In the bistable regions, there are two minima and one maximum that correspond to three different motional modes of the atoms in the third energy. As a consequence of the bistable curves or the loops structure, the existence of the multiple minima value at the fixed reduced detuning $B$ is a omnipresent sign of hysteresis \cite{Mueller}. When the detuning is larger and exceed the critical value $B_3$, the bistable curve emerges in the upper level, which corresponds to the loop structure of the upper level $E_1$ in Fig. \ref{Fig2}(e). The bistability can also be observed in the same ranges of detunings in the mean intracavity photon number $|\alpha|^2$, as shown in Fig. \ref{Fig3}(c). The results for  $\lambda=2$ cases are shown in Fig. \ref{Fig4}, with the increase of the detunings, the energy $E_2$ also shows bistability behavior between $B_1=-0.96$ and $B_3=-0.3$.

Figures \ref{Fig2}, \ref{Fig3}, and \ref{Fig4} focus on the non-interaction cases. Now we turn to the interacting BECs case. Figure \ref{Fig5} depicts the eigenenergy versus the zero-point energy bias $\lambda$ but covering different interaction strengths $r$. As we consider two types of the nonlinearity simultaneously, i.e., the atomic interaction and cavity-mediated nonlinearity. the energy structure is more complex than the one obtained for only one type of nonlinearity mentioned. It is shown that the structure is similar to the noninteraction case ($r=0$) when the parameter $r$ is weak ($r=1$). A loop structure appears at the middle level $E_2$ shown in Fig. \ref{Fig5}(b) as $r$ is increased ($r$=1). As $r$ increases to $1.5$ and $2.5$, respectively,  the loop structures have more complex shapes in the upper level $E_1$ and the middle level $E_2$ as shown in Figs. \ref{Fig5}(c) and \ref{Fig5}(d). Compared Fig. \ref{Fig5} with Fig. \ref{Fig2}, one can find that, this kind of complicated loop structure occurred in the upper and middle levels comes from the nonlinearity caused by the atom-atom interaction \cite{Liu1}. Furthermore,  the loop structure of the lower level $E_3$ does not change with the increase of the interatomic interaction as shown in Fig.\ref{Fig5}. Thus one can see that this loop structure in energy $E_3$ is mainly caused by the nonlinearity generated from the cavity back-action rather than from the direct interatomic interactions.

\begin{figure}[ptb]
\centerline{\includegraphics*[width=0.5\textwidth]{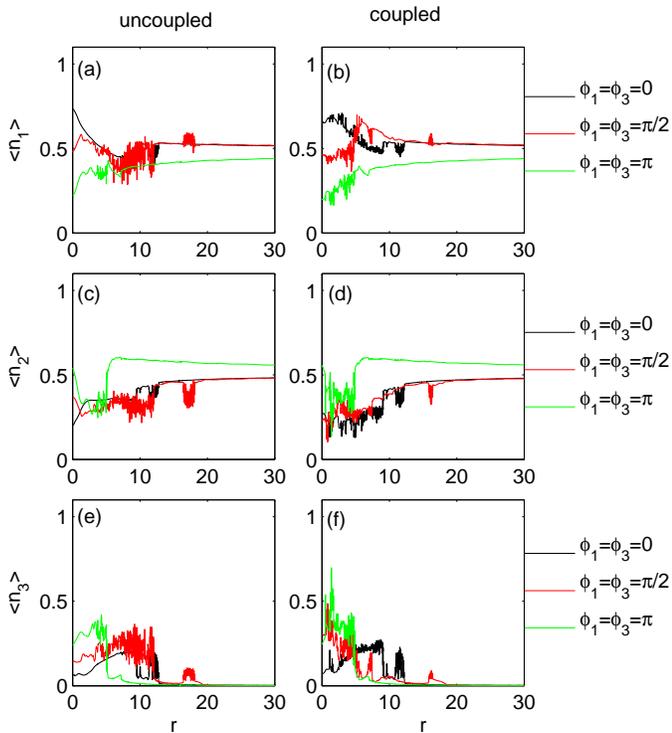}}\caption{(Color online) The average population of (a), (c), (e) (uncoupled case) and (b), (d), (f)(coupled case) $\langle n_i \rangle$  with the initial value $n_1=0.5$, $n_2=0.5$ and $n_3=0$ in the tilted wells $\lambda=-1$.  The other parameters of the cavity are chosen as $\delta U_0A^2/(2v)=0.02$, $B=-0.9$ and $C=0.07$.}%
\label{Fig11}%
\end{figure}

\begin{figure}[ptb]
\centerline{\includegraphics*[width=0.5\textwidth]{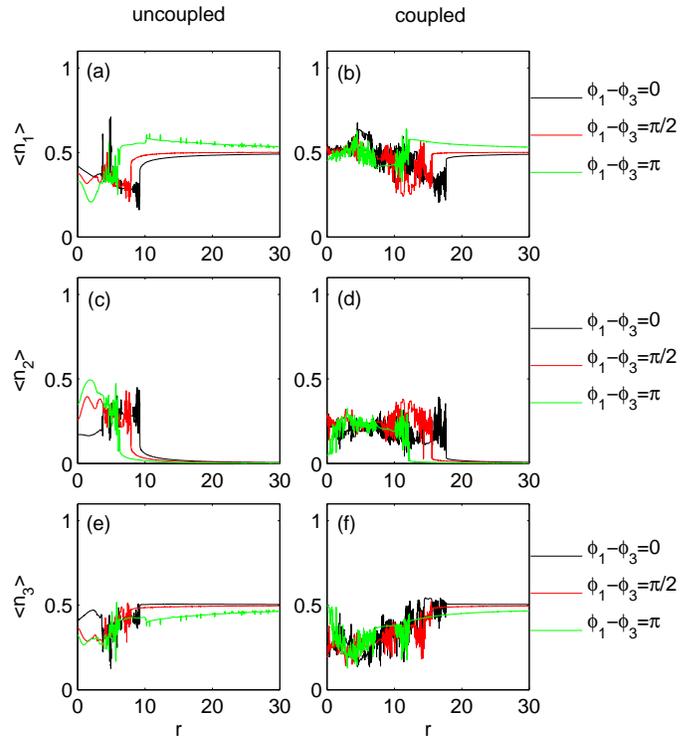}}\caption{(Color online) The average population of (a), (c), (e) (uncoupled case) and (b), (d), (f)(coupled case) $\langle n_i \rangle$  with the initial value $n_1=0.5$, $n_3=0.5$ and $n_2=0$ in the tilted wells $\lambda=-1$. The other parameters of the cavity are chosen as $\delta U_0A^2/(2v)=0.02$, $B=-0.9$ and $C=0.07$.}%
\label{Fig12}%
\end{figure}

\section{tunneling to localization, $\lambda=0$}
In this section, we will perform numerical simulation of localization or self trapping of a BEC in this cavity-mediated triple well system using the solution of the Eqs. (\ref{eq-7})-(\ref{eq-10}). In general, the localization or self trapping is caused by the nonlinear interaction between atoms, then the self trapping of a repulsive BECs system in a bare optical lattice, double well and triple well potential has been investigated firstly both for the dynamical and stationary cases \cite{Adhikari}. When considering the coupling to a cavity, the cavity-mediated atoms-field interaction will provide a novel nonlinearity different from the atoms-atoms nonlinear interaction, then the research shows the self-trapping behavior can be modified drastically by the strong condensate-field interplay \cite{Zhang,Zhang1,Larson2} in a double well potential.  As mentioned before, the cavity-mediated triple-well system induce a effective potential and tilt the symmetric well as a result.  The more asymmetric the well becomes, the more complex the dynamical behavior appears. Then it is interesting to investigate the localization and the self-trapping dynamics in this system. We will calculate the average population of each well for the different initial values with a different parameter $r$ in the following. As a benchmark, we firstly discuss the symmetry cases $\gamma=0$.

\subsection{localization in one well}
Comparing to a bare BECs in the triple-well, here the BECs in the cavity-mediated triple well systems couples to the cavity mode, the atom-cavity interplay terms, the last term of the Hamiltonian (\ref{eq-4}), will not only change the critical value $r_c$, but also alter the process of transition to localization. To illustrate this point, we calculate the mean value $\langle n_i \rangle$ ($i=1, 2, 3$) of the BECs uncoupled or coupled to the cavity mode for different interaction $r$ with the initial conditions $n_1(0)=1$ and $n_2(0)=1$, respectively.  The results are presented in Fig. \ref{Fig6}. When the BECs are uploaded initially in the left well, shown in Fig.\ref{Fig6}(b), it can be found that the average populations of each well are more chaotic than the ones of the uncoupled case (Fig. \ref{Fig6}(a)), which is due to the reasons that the nonlinearity of the effective atom-cavity induced potential in Eq. (\ref{eq-4}) tilts the wells and changes the stationary points of the condensates and makes some new motional modes of the atomic oscillation appear, correspondingly the critical value of the transition in the coupled case is larger than the uncoupled one. However, when the BECs are uploaded in the middle well initially, shown in Fig. \ref{Fig6}(c) and \ref{Fig6}(d),  the cavity-mediated nonlinearity does not have a pronounced effects on the transitions. It can be explained as follows: as mentioned before, the atom-cavity induced interaction is equivalent to the zero-energy $\lambda$, which affect the dynamics by tilting the symmetric well. When one put atoms in the middle well, the asymmetry of the wells $1$ and $3$ does not affect the dynamics of the BECs in well $2$ apparently. With the increase of the interaction, the oscillation will be compressed and the BECS atoms will be trapped in the middle well completely. Comparing the coupled cases with uncoupled cases, one can also find that the values of $\langle n_1 \rangle$ and $\langle n_2 \rangle$ are different for $r=0$, which is implied that the BECs atoms in the wells are rearranged when the condensates couple to the cavity.

\begin{figure}[ptb]
\centerline{\includegraphics*[width=0.5\textwidth]{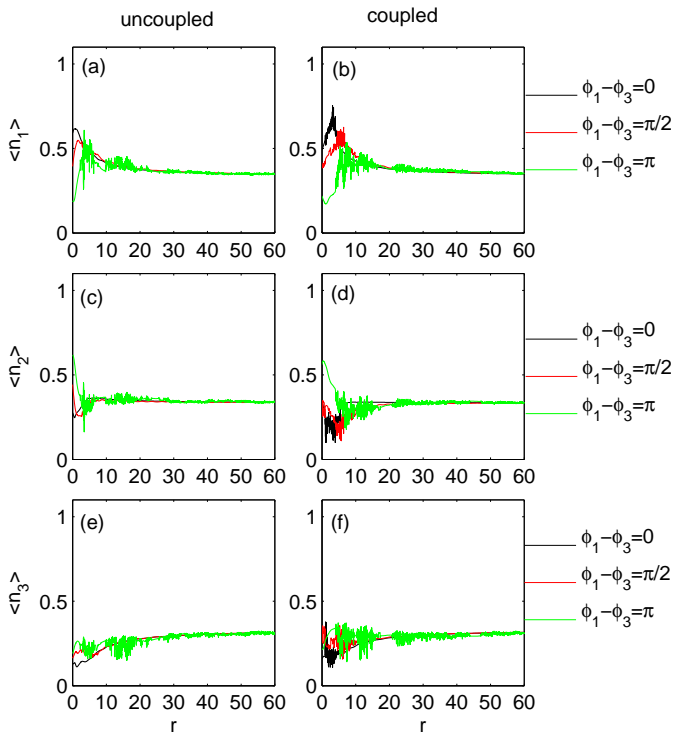}}\caption{(Color online) The average population of (a), (c), (e) (uncoupled case) and (b), (d), (f)(coupled case) $\langle n_i \rangle$  with the initial value $n_1=1/3$, $n_2=1/3$, $n_3=1/3$ for different relative phases in the tilted wells $\lambda=-1$. The other parameters of the cavity are chosen as $\delta U_0A^2/(2v)=0.02$, $B=-0.9$ and $C=0.07$.}%
\label{Fig13}%
\end{figure}
\subsection{localization in two- and three-wells}
This part will focus on the dynamics of the BECs atoms trapped initially in two- and three-wells with different initial relative phase. We calculate the average value $\langle n_i\rangle$ $(i=1,2,3)$ as a function of $r$ for different cases. $\langle n_i \rangle$ versus $r$ with the initial value $n_1=0.5$, $n_2=0.5$, and $n_3=0$ (the BECs atoms are initially uploaded in wells $1$ and $2$) is shown in Figs. \ref{Fig7}. For comparison, the uncoupled cases are also presented. As BECs are uploaded in two wells initially, the interference should be considered, and their initial phase will be important. In general, different initial phase will lead to different behaviors. From Fig. \ref{Fig7}, one can find that the coupling to the cavity will generate distinct effects for different relative initial phase. when the relative phase is zero, the atom-cavity coupling will make the smooth oscillation process become chaotic process for the small interaction $r$, as a result the coupling to the cavity will enlarge the chaotic region. While for the relative phase $\pi/2$ cases, this atom-cavity nonlinearity will decrease the chaotic region. With the increase of the interaction, the atoms will be trapped in wells $1$ and $2$ for the relative phases $\pi/2$ and zero, respectively. When the relative phase is set to $\pi$, compared Fig. \ref{Fig7}(f) with Fig. \ref{Fig7}(e), one can find that the atom will be trapped in wells $1$ and $2$ from small interaction $r$, but the atoms in each well oscillate quickly and the atoms can not be trapped easily in each well. Furthermore, the cavity interaction will change the population in wells $1$ and $2$. In the uncoupled cases, the population $\langle n_1\rangle$ is higher than $0.5$ for a large interaction region, the coupling to the cavity will lead the population $\langle n_1\rangle$ decrease to a lower value than $0.5$ for a large interaction region as shown in Fig.\ref{Fig7}(b). While for  $\langle n_2\rangle$,  the coupling to the cavity will lead the population $\langle n_2\rangle$ to be higher than $0.5$ for a large interaction region on the other contrary. Comparing Fig.\ref{Fig7}(d) with Fig. \ref{Fig6} (d), one can also find that the BECs in the middle well are more chaotic as atoms trapped in two wells initially due to the tunneling and interference coming from the BECs in the well $1$. The mean value $\langle n_i \rangle$ versus $r$  with the initial value $n_1=0.5$, $n_2=0$, and $n_3=0.5$ (the BECs atoms are initially uploaded in wells $1$ and $2$) is shown in Fig. \ref{Fig8}. It can be found that the atom-cavity coupling results into different effects for the different relative phase as mentioned in \ref{Fig7}. After passing a wide oscillation region, the oscillation and the chaos behavior are compressed into a narrow region and the atoms will be trapped equally for a large interaction $r$ in the two traps eventually where they are initially uploaded for three cases. From Fig.\ref{Fig8}(b), we can find that the noninteracting atoms are mostly trapped in well $1$ due to the coupling to the cavity for the relative phase $\pi/2$, but with the increase of the interaction, the population decrease quickly and enter a chaotic process. When the relative phase is chosen as zero, the cavity-mediated effects only enlarge the oscillation region, but for the other two relative phases cases, this nonlinearity not only enlarge the oscillation region but also make more chaotic behavior.  Then we can change the relative phases to control the dynamics of the BECs in this cavity-mediated triple well. The above are scheduled to study the localization of BECs in two wells. As the triple well system is a protype of optical lattices helping us provide a bottom-up understanding of mechanisms operating in the infinite optical lattice. Here, we will turn to the case that the BECs atoms are trapped in three wells simultaneously with initial $n_1=1/3$, $n_2=1/3$, and $n_3=1/3$. We calculate the average value $\langle n_i\rangle$ $(i=1,2,3)$ as a function of $r$ for different relative phases in Fig. \ref{Fig9}. The uncoupled cases are also shown for comparison. It can be easily found that the dynamics is also sensitively related to the initial phase but the phase does not show a pronounced effect as it does in the two wells cases. The coupling to the cavity also enlarges the oscillation region and make more chaotic for the cases of the relative phase $\pi/2$ and $\pi$. While the relative phase is zero, the atom-cavity interplay may be neglected for its very weak influence on the dynamics, the BECs atoms show same process moving smoothly to a robust trapped behavior from the a small interaction $r$ both for the coupled and uncoupled cases.

\section{tunneling to localization, $\lambda\neq0$}
In the above, we have studied the dynamics of the cavity-mediated BECs in the symmetric triple-well cases ($\lambda=0$). Though the cavity-mediated potential makes the triple-well tilted as the zero-energy  $\lambda$ does, the actual experiments values, $\delta U_0A^2/(2v)=0.02 << 1$ only leads the triple well to tilt slightly. How does the cavity-mediated potential affect the dynamics of the BECs for the asymmetric triple-well cases, and what will happen for a cavity-mediated tilted triple-well system?  In order to answer these interesting questions, this section will extended to the general case with $\lambda=-1$. In the following, the self trapping or localization in one-, two-, and three-wells are investigated, respectively.
\subsection{localization in one well}
 We explore the dynamics of of the BECs by depicting the mean populations for the uncoupled and coupled cases versus the parameter $r$ with $\lambda=-1$ and the initial conditions are $n_1(0)=1$, $n_2(0)=1$, and $n_3(0)=1$, respectively. The numerical results are shown in Figs. \ref{Fig10}(a)$-$\ref{Fig10}(f). In Fig.\ref{Fig10}(b), with an increase in the interaction $r$, the averaged population $\langle n_1 \rangle$ decreases smoothly at the beginning, passes a large oscillating and chaotic areas till $r=7.61$, and then tends to unite. Compared to the uncoupled case (Fig. \ref{Fig10}(a)), it can be found that the transition area and the critical value $r$ are larger for the coupled case due to the reason that the atom-cavity interplay enhances the oscillation of the BECs atoms of each wells. Thus a larger interaction $r$ is used to trap the condensates. Compared Fig.\ref{Fig10}(b) with Fig.\ref{Fig6}(b), one can also find that, when the interaction is small, the coupling to the cavity will help the BECs atoms nearly be trapped in well $1$ for the tilted case. In Fig. \ref{Fig10}(d), the cavity-mediated interaction will make the averaged population $\langle n_1 \rangle$ show a turbulent behavior from a small interaction $r$, passes a wide chaos region, and then jumps up and tends to unite. In Fig. \ref{Fig10}(f), initially, upload all the atoms in the right-hand well, with an increase in $r$, the average population $\langle n_3 \rangle$ undergo a smooth process, then tend to unite. In comparison to Fig. \ref{Fig10}(e), it is shown that the coupling to the cavity will lead to a larger critical value of the interaction $r$.

\subsection{localization in two- and three-wells}

Except the wells are tilted, the parameters and the initial conditions in Fig.\ref{Fig11} are as same as the ones in Fig.\ref{Fig7}. Compared with these two figures, we can find that besides the phase sensitivity and the enlarging chaotic region, the difference is, the condensates atoms for the relative phase $\pi$ will be trapped too for the tilted wells cases. Furthermore, the mean value of $\langle n_1 \rangle$ is always small than $\langle n_2 \rangle$ the due to the tilted wells, i.e., the atom-cavity induced tilt potential is weak than the tilt of the well, and then it can not make a inversion as it does in the symmetric well cases. The mean value $\langle n_i \rangle$ for different initial phases with the initial value $n_1=0.5$, $n_2=0$, $n_3=0.5$ for the tilted well cases is shown in Fig.\ref{Fig12}. Compared to Fig. \ref{Fig8}, we may find two characteristics due to the tilted well, one is the the smooth motion of atoms for the relative phase zero become more chaotic due to the common effects of the tilted well and the atom-cavity induced potential. The other is the trapped mean population of $\langle n_1 \rangle$ is large than $\langle n_1 \rangle$ due to the tilt. For comparison with Fig. \ref{Fig9}, the BECs atoms trapped in the tilted three wells simultaneously with initial $n_1=1/3$, $n_2=1/3$, and $n_3=1/3$ are also investigated shown in Fig.\ref{Fig13}. One obvious phenomena is for the relative phase zero case, when the well is tilted only, the mean values of $n_i$ is still smooth. When the tilted tripe well couples with the cavity,  the atom-cavity interaction will destroy this Smooth behavior and make the dynamics become chaotic for small interaction $r$.

\section{summary}
In this paper, we have studied the bistable curves and the tunneling dynamics of BECs in the cavity-mediated triple-well system within the mean field approach. We find that the atom-cavity  nonlinearity will make the bistability occur both in the number of the photons of the cavity and in the energy structure of the BECs. Furthermore, the loops appear for the upper and lower energy levels with the pump-cavity detuning increasing. We also demonstrated that the transition to localization of a BECs can be efficiently controlled by the optical cavity. To understand the effects resulting form the cavity coupling, we have performed a detailed numerical simulation both for the symmetric and asymmetric cases.

For the cases of BECs atoms uploaded in well $1$ initially, the atom-cavity field nonlinearity enlarges the oscillation and chaotic regions both for the symmetric and asymmetric cases, then a large interaction is needed to trap the atoms. When the BECs atoms are uploaded in well $2$ initially, this nonlinearity only have a pronounced effects for the asymmetric case, which makes more chaos.

For the cases of BECs atoms uploaded in two wells initially, the cavity have different effects for different relative phases.  With the increase of the interaction, the atoms will be trapped in one or two wells where they initially uploaded. Besides this, the coupling to cavity can lead a mean population inversion for certain relative phase if the BECs atoms are initially uploaded in wells $1$ and $2$.

For the cases of the BECs atoms uploaded in three wells initially, there is a robust trapped phenomena for the relative phase zero in the symmetric wells even for the coupled cases. The theoretical discussion here will be helpful in controlling of BECs in real experiments.

\begin{acknowledgments}
This work was supported by NSFC under grants Nos. 10704031,
10874235, 10934010 and 60978019, the NKBRSFC under grants Nos.
2009CB930701, 2010CB922904 and 2011CB921500, and FRFCU under grant
No. lzujbky-2010-75.
\end{acknowledgments}


\begin{thebibliography}{99}

\bibitem{Brennecke}F. Brennecke, T. Donner, S. Ritter, T. Bourdel, M. K\"{o}hl, and T. Esslinger, Nature (London) \textbf{450}, 268 (2007).
\bibitem{Colombe}Y. Colombe, T. Steinmetz, G. Dubois, F. Linke, D. Hunger,and J. Reichel, Nature (London) \textbf{450}, 272 (2007).
\bibitem{Cola}M. Cola, M. Paris, and N. Piovella, Phys. Rev. A \textbf{70}, 043809 (2004).
\bibitem{Slama}S. Slama, S. Bux, G. Krenz, C. Zimmermann, and P. W. Courteille, Phys. Rev. Lett. \textbf{98}, 053603 (2007).
\bibitem{Vukics}A. Vukics, C. Maschler, and H. Ritsch, New J. Phys. \textbf{9}, 255 (2007).
\bibitem{Maschler}C. Maschler, I. B. Mekhov, and H. Ritsch, Eur. Phys. J. D \textbf{46}, 545 (2008).
\bibitem{Zhang}J. M. Zhang, W. M. Liu, and D. L. Zhou, Phys. Rev. A \textbf{77}, 033620 (2008).
\bibitem{Larson}J. Larson, B. Damski, G. Morigi, and M. Lewenstein, Phys. Rev. Lett. \textbf{100}, 050401 (2008).
\bibitem{Szirmai}G. Szirmai, D. Nagy, and P. Domokos, Phys. Rev. Lett. \textbf{102}, 080401 (2009).

\bibitem{Ruostekoski} U. Shrestha and J. Ruostekoski, New J. Phys. \textbf{14}, 043037 (2012).


\bibitem{Murch}K. W. Murch, K. L. Moore, S. Gupta, and D. M. Stamper-Kurn, Nature Phys. \textbf{4}, 561 (2008).
\bibitem{Brennecke1}F. Brennecke, S. Ritter, T. Donner, and T. Esslinger, Science \textbf{322}, 235 (2008).
\bibitem{Baumann}K. Baumann, C. Guerlin, F. Brennecke, and T. Esslinger, Nature (London) \textbf{464}, 1301 (2010);


\bibitem{Zhou}L. Zhou, H. Pu, H. Y. Ling, and W. P. Zhang, Phys. Rev. Lett. \textbf{103}, 160403 (2009); L. Zhou, H. Pu, H. Y. Ling, K. Zhang, and W. P. Zhang, Phys. Rev. A \textbf{81}, 063641 (2010).
\bibitem{Larson1}J. Larson and M. Lewenstein, New J. Phys. \textbf{11}, 063027 (2009).
\bibitem{Dong}Y. Dong, J. W. Ye, and H. Pu, Phys. Rev. A \textbf{83}, 031608 (2011).


\bibitem{Gupta}S. Gupta, K. L. Moore, K. W. Murch, and D. M. Stamper-Kurn, Phys. Rev. Lett. \textbf{99}, 213601 (2007).
\bibitem{Ritter}S. Ritter, F. Brennecke, K. Baumann, T. Donner, C. Guerlin, T. Esslinger, Appl. Phys. B \textbf{95}, 213 (2009).
\bibitem{Szirmai1}G. Szirmai, D. Nagy, and P. Domokos, Phys. Rev. A \textbf{81}, 043639 (2010).
\bibitem{Yang} S. Yang, M. Al-Amri, J. Evers, and M. S. Zubairy, Phys. Rev. A \textbf{83}, 053821 (2011).
\bibitem{Prasanna}B. P. Venkatesh, J. Larson, and D. H. J. O'Dell, Phys. Rev. A \textbf{83}, 063606 (2011).
\bibitem{Zhou1} L. Zhou, H. Pu, K. Zhang, X. D. Zhao, and W. P. Zhang, Phys. Rev. A \textbf{84}, 043606 (2011).

\bibitem{Keeling}J. Keeling, M. J. Bhaseen, and B. D. Simons, Phys. Rev. Lett. \textbf{105}, 043001 (2010).

\bibitem{Nagy}D. Nagy, G. K\'{o}nya, G. Szirmai, and P. Domokos, Phys. Rev. Lett. \textbf{104}, 130401(2010).
\bibitem{Liu}N. Liu, J. L. Lian, J. Ma, L. T. Xiao, G. Chen, J. Q. Liang, and S. T Jia, Phys. Rev. A \textbf{83}, 033601(2011)

\bibitem{Zhang1}J. M. Zhang, W. M. Liu, and D. L. Zhou, Phys. Rev. A \textbf{78}, 043618 (2008).
\bibitem{Larson2}J. Larson, J. Martikainen, Phys. Rev. A \textbf{82}, 033606 (2010).
\bibitem{Chen}W. Chen and P. Meystre, Phys. Rev. A \textbf{79}, 043801(2009).

\bibitem{Mossmann}S. Mossmann and C. Jung, Phys. Rev. A \textbf{74}, 033601 (2006).
\bibitem{Lahaye}T. Lahaye, T. Pfau, and L. Santos, Phys. Rev. Lett. \textbf{104}, 170404 (2010).
\bibitem{Liu1} B. Liu, L. B. Fu, S. P. Yang, and J. Liu,  Phys. Rev. A  \textbf{75}, 033601 (2007).
\bibitem{Cao}L. Cao, I. Brouzos, S. Z\"{o}lner and P. Schmelcher, New J. Phys. \textbf{13}, 033032 (2011).

\bibitem{Graefe}E. M. Graefe, H. J. Korsch, and D. Witthaut, Phys. Rev. A \textbf{73}, 013617 (2006).
\bibitem{Lu}G. B. Lu, W. H. Hai, and Q. T. Xie, Phys. Rev. A \textbf{83}, 013407 (2011).
\bibitem{Streltsov}A. I. Streltsov, K. Sakmann, O. E. Alon, and L. S. Cederbaum, Phys. Rev. A \textbf{83}, 043604 (2011).
\bibitem{Stickney} J. A. Stickney, D. Z. Anderson, and A. A. Zozulya, Phys. Rev. A \textbf{75}, 013608 (2007).
\bibitem{Benseny} A. Benseny, S. Fern\'{a}ndez-Vidal, J. Bagud\`{a}, R. Corbal\'{a}n, A. Pic\'{o}n, L. Roso, G. Birkl, and J. Mompart, Phys. Rev. A \textbf{82}, 013604 (2010).

\bibitem{Haroche} S. Haroche, M. Brune, and J. M. Raimond, Europhys. Lett. \textbf{14}, 19 (1991)

\bibitem{Parameters} The set of parameters is experimentally practical. By taking $\kappa$ $\sim$ $2\pi \times 10^6$ Hz, $U_0$ $\sim$ $2\pi \times 10^4$, N $\sim$ $10^4$ and $A$ $\sim$ $10^{-3}$, we have $\delta U_0A^2/(2v)$ $\sim$ 0.01, $C$ $\sim$ 0.01, and $B$ is changed according to the pump-cavity detuning.

\bibitem{Wu} B. Wu and Q. Niu, Phys. Rev. A \textbf{64}, 061603 (2001); B. Wu and Q. Niu, New J. Phys. \textbf{5}, 104 (2003).
\bibitem{Diakonov}D. Diakonov, L. M. Jensen, C. J. Pethick, and H. Smith, Phys. Rev. A \textbf{66}, 013604 (2002); M. Machholm, C. J. Pethick, and H. Smith, ibid. \textbf{67}, 053613 (2003); M. Machholm, A. Nicolin,C. J. Pethick, and H. Smith, ibid. \textbf{69}, 043604 (2004).
\bibitem{Watanabe}G. Watanabe, S. Yoon, and F. Dalfovo, Phys. Rev. Lett. \textbf{107}, 270404 (2011).
\bibitem{Watanabe1}G. Watanabe, H. Sonoda, T. Maruyama, K. Sato, K. Yasuoka, and T. Ebisuzaki, Phys. Rev. Lett. \textbf{103}, 121101 (2009).

\bibitem{Mueller} E. J. Mueller, Phys. Rev. A \textbf{66}, 063603 (2002).

\bibitem{Adhikari}S. K. Adhikari, J. Phys. B \textbf{44}, 075301 (2011).


\end{thebibliography}
\end {document}